\documentclass[prb,twocolumn,showpacs,floatfix]{revtex4}%
\usepackage{graphicx}% Include figure files
\usepackage{dcolumn}% Align table columns on decimal point
\usepackage{bm}% bold math
\usepackage{latexsym}% bold math

\begin{document}

\title{ Size driven phase transitions in pinned vortex systems}
\author{P. S. Cornaglia}
\author{M. F. Laguna}
\affiliation{Centro At\'{o}mico Bariloche, 8400 San Carlos 
de Bariloche, R\'{\i}o Negro, Argentina}

\date{\today}

\begin{abstract}
We model a tridimensional vortex system in a sample with square superficial pinning in the top surface and obtain the ground state structures as a function of the sample thickness. Using a simple Frenkel-Kontorova like model and no adjustable parameters, we reproduce the experimental vortex configurations seen in the bottom surface and their range of stability. We find three phases with two transitions between them, including a continuous one from square to distorted hexagonal structure and a discontinuous one from distorted hexagonal to hexagonal structure.

\end{abstract}


\maketitle

In recent years the phase diagram of vortices in high temperature superconductors has been the subject of intense experimental and theoretical studies.\cite{crabtree} In Bi$_2$Sr$_2$CaCu$_2$O$_8$ crystals the vortex liquid transforms, upon cooling, for low enough magnetic fields into a Bragg Glass through a first-order phase transition. The Bragg Glass has hexagonal quasi-long range order and is collectively pinned.\cite{Giamarchi} The presence of uncorrelated pinning potentials,\cite{Blatter} as well as periodic pinning potentials,\cite{Hoffmann, Laguna} leads to the appearance of a rich variety of phases and transitions between them. The static and dynamic properties of vortex systems with periodic pinning arrays have been extensively studied experimentally\cite{Baert, Harada} and, for two-dimensional systems, numerically.\cite{Laguna, Reichhardt, Reichhardt2, Marconi}

Superficial pinning centers with different spacial distributions can be artificially created,\cite{Fasano} and the magnetic decoration technique allows the study of the vortex solid topology on the sample surface with a single-vortex resolution.\cite{Essmann, Fasano2} In recent experiments, the structure of the vortex system in the presence of a superficial pinning potential has been studied in Bi$_2$Sr$_2$CaCu$_2$O$_8$ single crystals.\cite{Fasano_sq} On the sample top surface, a square array of Fe dots is created by means of electron-beam lithography. These dots act as pinning centers for the vortices and are strong enough to determine their positions at the top surface in the ground state. Performing magnetic decorations on the opposite surface of the sample, with the magnetic field perpendicular to it, different vortex structures as a function of the sample thickness are observed. For a vortex density equal to the pinning centers density and thick enough samples the natural hexagonal structure is observed. This implies that the vortices deform themselves along the z-axis to connect the square and hexagonal lattices. For thin samples a square vortex array is found, while for intermediate sample thickness, a polycrystal with grains of distorted hexagonal and square structure is obtained.

Despite the theoretical work done on the behavior of vortices in periodic potentials, to our knowledge, the case of the three-dimensional vortex system with two-dimensional pinning has not been addressed.
Note that a direct continuous transition between square and triangular lattices, as described for example by Landau theory, cannot occur by symmetry arguments. Therefore, there are no obvious expectations on the evolution of the vortex structure with thickness.

In this work we present a model for a three-dimensional vortex system in interaction with a two-dimensional pinning array at one surface of the sample and study the vortex configurations as a function of the sample thickness $L_z$.
We consider the case of square superficial pinning at the first matching field (equal number of vortices and pinning centers) and study the strong pinning case of Ref.[14], where the vortices in the ground state are pinned at one end. There are two competing energies, the length energy and the vortex-vortex interaction energy.  On one hand, the minimum length energy is obtained for parallel straight vortices of length $L_z$. Since the vortices are fixed at one end to the square lattice of the top surface, this energy is minimized when the vortex structure is square at the bottom. On the other hand, the vortex-vortex interaction favors a triangular structure.

For vortices tilted at small angles the vortex-vortex interaction potential can be approximated by:\cite{Koshelev,Blatter,Buzdin}
\begin{equation}\label{int}
V\left[{\mathbf u_i}(z),{\mathbf u_j}(z)\right] =\frac{L_z}{2}\left\{ V_s\left[ u_{ij}(0)\right]+ V_s\left[ u_{ij}(L_z)\right]\right\}.
\end{equation}
Here $ u_{ij}(z)=|{\mathbf u_{i}}(z)-{\mathbf u_{j}}(z)|$, where ${\mathbf u_{i}}(z)=(x_i(z),y_i(z))$ is the planar coordinate of the {\it i}th vortex and

\begin{equation}\label{interaction} 
V_{s}\left(u\right) =2\epsilon_0 \left[ -\log\left| u \right| +  u^2 -  \frac{ u^4}{4} -\frac{3}{4} \right].
\end{equation}
This expression has a cut-off at $u_c=2.5\lambda$, where $\lambda$ is the penetration depth; $u$ is in units of $u_c$ and $\epsilon_0 = \left(\phi_0/{4 \pi \lambda}\right)^2$. We checked that using the computationally more costly modified Bessel function (exact for parallel rigid vortices) instead of (\ref{interaction}) gives very similar results.
 
\begin{figure}[tbp]
\includegraphics[width=7cm,clip=true]{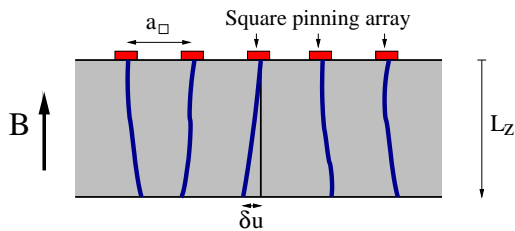}
\caption{ Schematic section of the sample parallel to a line of pinning centers and to the magnetic field B. The vortex projections, perpendicular to the plane of the figure, are drawn with one of their ends pinned by the superficial pinning potential.}
\label{fig1}
\end{figure}

We write the length energy of the vortex $i$ as a function of the relative displacement of the vortex ends $\delta u_i=\left|{\mathbf u_i}(L_z)-{\mathbf u_i}(0)\right|$ (see Fig. \ref{fig1}). Since the $\delta u_i/L_z$ are very small, we can make a Taylor series expansion. The zero-th order term, $e^0_l=L_z \epsilon_0 \ln \kappa$, is the energy of a straight vortex, where $\kappa$ is the Ginzburg-Landau parameter. The next non zero contribution to the length energy is $\frac{1}{2}\epsilon_0 \left(\delta u_i/L_z\right)^2\ln\kappa$. Since the results are not critically dependent on higher order terms, in this work we used the following expression:
\begin{equation}\label{length} 
e_{l}\left({\delta u_i}\right) =  \epsilon_0 \ln \kappa \sqrt{{{\delta u_i}}^2+L_z^2},
\end{equation}
that corresponds to the length energy of a vortex tilted an angle $\arctan\left(\delta u_i/L_z\right)$. We model the effect of the surface pinning keeping the vortex top coordinates ${\mathbf u}_i(0)$ fixed to a square lattice of lattice parameter $a_\Box$.

The total energy of the vortex system is the sum of the interaction energy over all the vortex pairs, and the length energy over all the vortices. We shift this energy adding a constant term,  $-\frac{1}{2}(E_\Box+E_\triangle)-N e_l^0$, where $E_\triangle$ and $E_\Box =L_z\sum_{i\neq j} V_s\left[u_{ij}(0)\right]$ are the interaction energies for a triangular and square array of straight parallel vortices respectively, and $N$ is the total number of vortices:
\begin{equation}\label{hamiltonian} 
E_{tot}= \frac{L_z}{2}\sum_{i\neq j} V_s\left[u_{ij}(L_z)\right]-\frac{E_\triangle}{2} + \sum_i^N \left[e_{l}\left({\delta u_i}\right)-e_l^0\right].
\end{equation}
This is a Frenkel-Kontorova like model in which the first two terms correspond to the interaction between adsorbed atoms (strain energy) and the last term models the effect of the periodic substrate (elastic energy).
Similar models\cite{Chaikin} have been used to study phase transitions of overlayers on surfaces.\cite{Hamilton} The results of this paper can be applied to these systems, where the decrease in the elastic energy --given by the substrate potential intensity-- can be related to an increase in the number of overlayers.\cite{Hamilton1}

We generated vortex structures changing slightly the primitive lattice vectors of the triangular lattice in such a way that the vortex lattice is commensurate with both pinning potential and sample size. To obtain the ground state configurations, we performed a simulated annealing down to zero-temperature using a hybrid algorithm. At each time step, we update the vortex positions using Langevin dynamics.\cite{Laguna2002} To improve the convergence we allow vortex reconnections by means of a Metropolis algorithm.\cite{Binder} Every five time-steps, two neighboring vortices ($i$ and $j$) can swap their free ends ${\mathbf u}_i(L_z)$ and ${\mathbf u}_j(L_z)$, with a probability that depends on the energy difference between configurations.

For all the ground state configurations we checked that the main assumption of our model remains valid, i.e. that $\delta u_i/L_z \ll 1$. Moreover, for finite and small enough temperatures this is also the case. The highest temperature that can be studied with this model grows with decreasing $L_z$ and reaches its maximum at $L_z=0$.
The number of vortices $N$ is in the range $[1024, 2548]$ and we use periodic 
boundary conditions in $x$ and $y$ directions. Here we present results for a magnetic field of $37$ Gauss ($a_\Box = 0.74\mu m$) and $\kappa=100$. 
Since the disorder freezes the vortex structure at temperatures $T \lesssim T_m $,\cite{Laguna2002} we have used the penetration length at the melting temperature $\lambda(T=T_m)=0.8\mu m$ rather than $\lambda(T=0)$, to compare with the experimental results. 

Hereafter we focus on the geometric structure of the free ends of the vortices ${\mathbf u}_i(L_z)$, that can be observed experimentally by means of the magnetic decoration technique.\cite{Fasano_sq}

A free square vortex lattice is unstable under the displacement of a vortex line along the (10) or (01) lattice directions. Moreover, for a given displacement $\delta u$, the system decreases further its energy moving neighboring lines of vortices in opposite directions. In an analogous way, when the vortices are held fixed at one end on a square lattice by means of a strong superficial pinning, they can decrease their interaction energy tilting themselves. In this case, there is also an increase in the length of the vortices with an associated increase in the length energy. If the length energy increase rate, as a function of the tilting angle, is greater than the interaction energy decrease rate, the square lattice with all the vortices straight is stable. This is the case for small samples where the length energy dominates over the interaction energy. 
\begin{figure}[btp]
\includegraphics[width=7cm,clip=true]{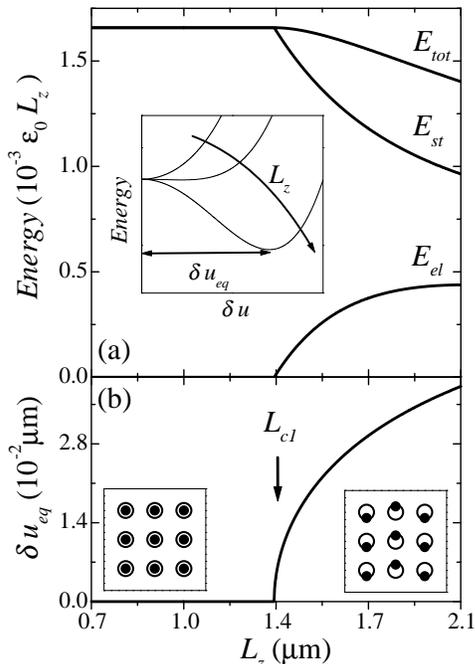}
\vspace{-5mm}
\caption{(a) Total ($E_{tot}$), strain ($E_{st}$) and elastic ($E_{el}$) ground state energies per vortex. Inset: $E_{tot}$ vs. $\delta u$ for different $L_z$; the minimum determines $\delta u_{eq}$  (b) Equilibrium vortex displacement $\delta u_{eq}$ as a function of the sample thickness $L_z$. A sharp increase at $L_{c1} \sim 1.4 \mu m$ is observed and corresponds to the transition from square to distorted hexagonal. Insets: square (left) and distorted hexagonal (right) structures. Open symbols correspond to pinning sites (top surface) and filled symbols to the free ends of the vortices (bottom surface).}
\label{fig2}
\end{figure}

In Fig. \ref{fig2}(a) we plot the ground state energies as a function of the sample thickness. For $L_z\le L_{c1}\sim1.4 \mu m$ the elastic energy is zero and the vortices are straight in a square lattice. For $L_z > L_{c1}$ the system can decrease its total energy tilting the vortices. This is associated to an increase of the elastic energy and to a decrease of the strain energy. The ground state geometry changes at $L_{c1}$ from square to distorted hexagonal. The distorted hexagonal structure is obtained tilting lines of vortices as shown in the insets of Fig. \ref{fig2}(b). This structure is fourfold degenerate (the other three structures can be obtained by $n \pi / 2$ rotations with $n$ integer) and can be described with a single parameter $\delta u_{eq}=\delta u_i$ (in this structure the $\delta u_i$ values are the same for all vortices). 
For $L_z < L_{c1}$ the total energy as a function of the vortex displacement $\delta u$ has a single minimum at $\delta u = 0$, as shown in the inset of Fig. \ref{fig2}(a). For $L_z > L_{c1}$ the $\delta u = 0$ state becomes unstable,  $\delta u_{eq}$ increases monotonously from zero and the strain energy decreases. If we keep this geometry fixed and let $L_z \to \infty$, the vortex displacements $\delta u_{eq}$ reach their maximum value $a_{\Box}/3$.  For vortex displacements along the $\hat{y}$ direction, we obtain an hexagonal structure compressed in the $\hat{x}$ direction by a factor $a_{{\Box}}/a_{\triangle}$ and expanded in the $\hat{y}$ direction by the inverse factor. Here $a_\triangle$ is the lattice parameter of the triangular structure and satisfies $a_\triangle^2 = 2 a_\Box^2/\sqrt{3}$.

\begin{figure}[tbp]
\includegraphics[width=7.5cm,clip=true]{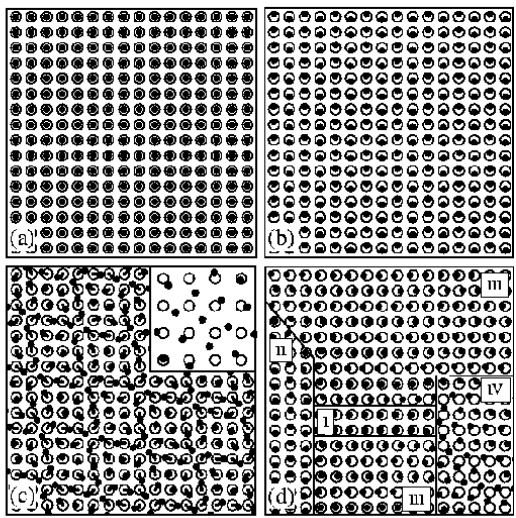}
\caption{Typical structures for the free ends (filled symbols) of the vortices in the three $L_z$ regimes. Open symbols correspond to pinning centers positions. (a) Square. (b) Distorted hexagonal. (c) Hexagonal, here the full vortex projection is also shown. Inset: detail of the structure. (d) Metastable state, obtained for $L_z = 4.1\mu m$, with grains of different structure: I) square, II) and III) distorted hexagonal in different orientations and IV) hexagonal.}
\label{fig3}
\end{figure}
The square and distorted hexagonal structures are shown in Fig. \ref{fig3}(a) and Fig. \ref{fig3}(b) respectively. For the distorted hexagonal structure we show one of the four possible configurations. 
For $L_z$ large enough the distorted hexagonal structure becomes metastable. In fact, the minimum energy configuration for $L_{z} > L_{c2}$ corresponds to a rotated hexagonal structure, slightly distorted to fit the pinning geometry (inset of Fig. \ref{fig3}(c)), that for simplicity we will call hexagonal. This structure can be generated with the lattice vectors $\vec{a}_1 = \alpha_1 a_\triangle \hat{x}$ and $\vec{a}_2 = (\alpha_1/2) a_\triangle \hat{x} +(\sqrt{3}\alpha_2/2) a_\triangle \hat{y}$, where $\alpha_1=3^{5/4}/4\sim 0.99$ and $\alpha_2=1/\alpha_1$, and it is commensurate with the square lattice when it is rotated $45$ degrees. The large $L_z$ regime was also studied using a model with more internal degrees of freedom per vortex, in which the sample is divided in equal thickness layers along the $\hat z$ direction. Each vortex is composed of segments with an interaction in each layer given by (\ref{interaction}) and length energy relative to adjacent layers given by (\ref{length}). In the range of studied $L_z$ values we find no significant difference between both methods. As can be seen in Fig. \ref{fig3}(c), where we plot the segment projections, the vortices deviate along the $\hat{z}$ axis only slightly from a straight line joining their ends at the sample surfaces.

It is interesting to analyze how the symmetry breaking occurs to reach the structure of Fig. \ref{fig3}(b) from high temperatures. This problem is related to the problem of melting in two dimensions over a square substrate that has been studied both theoretically and experimentally.\cite{Nelson1,Hamilton1,Muller1996} In these systems, the substrate acts as an Ising perturbation for the orientational order of the overlayer, and for smooth substrates an Ising-like phase transition is expected as a function of temperature.\cite{Nelson1}
For our vortex system with sample thicknesses $L_z \gtrsim L_{c1}$ (a ``strong substrate'' situation) we observe a related behavior at finite temperatures. We find a temperature regime in which several grains with different structures are observed: distorted hexagonal in its four possible orientations and square. As the temperature decreases, the grains merge into a single crystal of distorted hexagonal structure (Fig. \ref{fig3}(b)). However, if the vortex system is quenched, it gets trapped in a local energy minimum with a mixture of structures, as shown in Fig. \ref{fig3}(d). In the experiments of Ref. [14] the random pinning present in the samples produces an effective slowing down in the dynamics\cite{Laguna2002} and prevents the vortex system from reaching the ground state. For the range of $L_z$ where we find a distorted hexagonal structure, the experimental final states are a mixture of distorted hexagonal, square and hexagonal structures. In this regime there are several degenerate structures and the energy of a grain boundary between them vanishes at $L_z=L_{c1}$. As a consequence, for finite disorder, a polycrystalline structure is expected with its grains decreasing in size as we approach $L_{c1}$ from above.

\begin{figure}[tbp]
\includegraphics[width=8cm,clip=true]{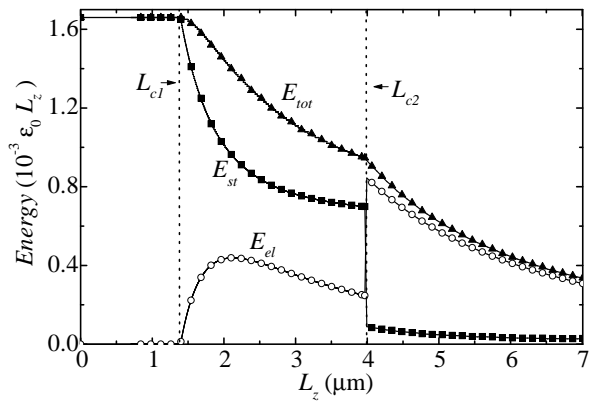}
\vspace{-2mm}
\caption{Ground state energies per vortex as a function of the sample thickness. There are three different phases separated by vertical bars at $L_{z}= L_{c1}$ and $L_{z}= L_{c2}$ . From left to right: square, distorted hexagonal and hexagonal phases.}
\label{fig4}
\end{figure}

At the transition from distorted hexagonal to hexagonal, there is a discontinuity in the elastic and strain energies and in the first derivative of the total energy, as can be observed for $L_z=L_{c2}$ in Fig. \ref{fig4}. In this case, there is no continuous way to go from one structure to another through minimum energy configurations. 

As $L_z$ is further increased we expect, in analogy with 1D systems, a series of transitions between states with decreasing strain energies or eventually a commensurate-incommensurate transition. However, the strain energy difference between hexagonal lattices of different commensurations and rotation angles is very small and we expect the orientation and deformations to be determined by the random pinning, instead of the superficial pinning geometry.

We showed, for the first matching field, that the system can be mapped to a 2D Frenkel-Kontorova model from the known vortex-vortex interactions and experimental parameters. With this simple model, we have reproduced the experimental results for the vortex structure as a function of the sample thickness. 
We predict two solid-solid phase transitions at $L_{z} = L_{c1}$ and $L_z=L_{c2}$. The first one, from square to distorted hexagonal, is continuous and can be described by a single parameter $\delta u$. The second one is first order and corresponds to the transformation from distorted hexagonal to hexagonal. To check our results we also analyzed a model with more internal degrees of freedom per vortex. This last model is appropriate to study, in the large $L_z$ case, how is the square to triangle transition along the $\hat{z}$ direction, a problem that will be addressed elsewhere.

The possibility of tunning the different parameters in the vortex systems allows the study of Frenkel-Kontorova models in the presence of disorder in a controllable way. This systems present a rich variety of phases and transitions between them. 

We thank F. de la Cruz, E. Jagla, Y. Fasano, M. De Seta, M. Menghini, A.A. Aligia, D. Dom\'{\i}nguez, A.B. Kolton, and C.A. Balseiro for useful discussions, and CONICET for financial support.

\end{document}